\documentstyle[pra,aps,amsfonts,amssymb,graphics,epsfig,color]{revtex}

\begin{document}

\title{Quadrature--dependent Bogoliubov transformations \\
       and multiphoton squeezed states}

\vspace{0.8cm}

\author{{\large Silvio De Siena, Antonio Di Lisi, 
and Fabrizio Illuminati}}

\address{
Dipartimento di Fisica, Universit\`a di Salerno, INFM -- Unit\`a 
di Salerno, and \\
INFN -- Sez. di Napoli, Gruppo coll. di Salerno, I--84081 Baronissi 
(SA), Italy \\
e--mail: desiena@sa.infn.it, dilisi@sa.infn.it, illuminati@sa.infn.it}

\date{\today}

\maketitle

\begin{abstract}
We introduce a linear, canonical transformation of the fundamental
single--mode field operators $a$ and $a^{\dagger}$ that generalizes 
the linear Bogoliubov transformation familiar in the construction of
the harmonic oscillator squeezed states.
This generalization is obtained by adding to the linear transformation 
a nonlinear function of any 
of the fundamental quadrature operators $X_{1}$ and $X_{2}$,
making the original Bogoliubov transformation quadrature--dependent.
Remarkably, the conditions of canonicity do not impose any 
constraint on the form of the nonlinear function, 
and lead to a set of nontrivial algebraic relations between 
the $c$--number coefficients of the transformation.
We examine in detail the structure and the properties of the 
new quantum states defined as eigenvectors of the transformed 
annihilation operator $b$. These eigenvectors
define a class of multiphoton squeezed states. 
The structure of the uncertainty 
products and of the quasiprobability distributions in phase space
shows that besides coherence properties, these
states exhibit a squeezing and a deformation 
(cooling) of the phase--space trajectories, 
both of which strongly depend
on the form of the nonlinear function.
The presence of the extra nonlinear term in
the phase of the wave functions 
has also relevant consequences on photon statistics
and correlation properties.
The non quadratic structure of the associated Hamiltonians
suggests that these states be generated
in connection with multiphoton processes in media with 
higher nonlinearities.
\end{abstract}

\vspace{0.2cm}


\vspace{0.2cm}

\section{Introduction}

The generalization of the squeezed states of the
harmonic oscillator to different systems and the
definition of new nonclassical states of light
have been actively pursued
in recent years \cite{sandberg}, \cite{eberly}, \cite{hillery}
\cite{nieto1}, \cite{nieto2}, \cite{prakash}, \cite{marian}. 
The possibility
of experimental realization has also been discussed
in some instances \cite{agarwal}, \cite{manko}, \cite{dematos}. 
However, 
these proposed generalizations cannot be easily
and straightforwardly connected to canonical and 
Hamiltonian structures.
Generalizations providing
a Hamiltonian structure and at the same
time preserving the canonical commutation relations
would thus be extremely
desirable not only from a mathematical and
conceptual point of
view, but expecially for possible connections
to new physical systems and experimental realizations.

In this paper we present a natural
generalization
of the Bogoliubov transformation originally 
introduced by Yuen to define the two--photon
coherent states of the harmonic oscillator
\cite{yuen}. This generalization
is achieved by simply
adding to the original transformation, which
is linear in the fundamental mode variables
$a$ and $a^{\dagger}$, a nonlinear operator--valued
function $F(X_i)$ of any of the fundamental quadratures
$X_1$ and $X_2$. The nonlinear function $F$ is by definition
hermitian, and thus defines a quadrature--dependent
Bogoliubov transformation. 
The transformed modes $b$ and $b^\dagger$ obey 
the canonical commutation relations $[b,b^\dagger]$
under suitable constraints on the coefficients
of the transformation. Such constraints are in
fact broad enough to allow for enough freedom
on the variation of the coefficients.
The eigenvectors of the transformed annihilation
operator $b$ define the quadrature--dependent
generalization of the squeezed states of the
harmonic oscillator. Such states are thus
multiphoton squeezed states associated to non quadratic
Hamiltonian systems. They preserve some coherence
properties, such as the classical motion of
the wave--packet center, but in addition they 
exhibit nonclassical features such as amplitude
squeezing, interference and deformation in
phase space, super--Poissonian modifications
of the field statistics and of the second--order
correlation functions. To avoid possible sources
of confusion, we emphasize that the multiphoton
squeezed states defined in the present work should
not be confused with the nonlinear coherent states
defined in the literature \cite{dematos}
as the eigenstates
of nonlinear operator valued functions of the number 
operator. In fact, both the physical origin and the 
mathematical properties of these two classes of states 
are rather different; they are both associated to
multiphoton down conversion processes, but the 
multiphoton squeezed states are generated in multiphoton
down conversion processes in optical parametric 
amplifiers of higher $\chi$ nonlinearity, while
the nonlinear coherent states are produced in
multiphoton down conversion processes in the
quantized motion of a trapped atom (the counterpart
of this process has also been recently studied
\cite{wallentowitz}).

The plan of the paper is the following:
the quadrature--dependent Bogoliubov transformation
is introduced in Sec. II, and the conditions of
canonicity are discussed. In Sec. III we introduce
and discuss in some detail the Hamiltonian structure 
induced by the transformation; in particular, we
provide the explicit form of the Hamiltonian expressed
in terms of the original mode variables for the lowest
nonlinear function of the amplitude $F(X_1)=X_1^2$.
The properties of the
uncertainty products are studied in Sec. IV, where 
it is shown that the multiphoton squeezed states
associated to the quadrature--dependent Bogoliubov
transformation exhibit amplitude squeezing and are
moreover asymptotically of minimum uncertainty.
The eigenvalue equation defining the multiphoton
squeezed states are explicitly solved in the
quadrature representation in Sec. V, and the 
coherent motion
of the wave--packet center is obtained.
Results for the photon number distributions and
second--order correlation functions are presented
in Sec. VI, while in Sec. VII we perform the 
phase--space analysis of the multiphoton squeezed
states in terms of the Wigner quasiprobability 
distributions and of the classical parametric
trajectories. Comments and conclusions follow
in Sec. VIII.

\section{The Quadrature--dependent Bogoliubov Transformation} 

Let us consider a single--mode bosonic quantum system.
The fundamental adimensional variables are the
bosonic annihilation and creation operators
$a$ and $a^{\dagger}$, with canonical commutation 
relation $[a,a^{\dagger}]=1$.

We introduce a new set of fundamental variables
$b$ and $b^{\dagger}$ by the linear transformation
\begin{equation}
b = \mu a + \nu a^{\dagger} 
+ \gamma F(X_1) + \eta G(X_2) \; ,
\label{eq:Equaz1}
\end{equation}

\noindent and its adjoint
\begin{equation}
b^{\dagger} = \mu^{*} a^{\dagger} + \nu^{*} a
+ \gamma^{*} F(X_1) + \eta^* G(X_2) \; .
\label{eq:Equaz2}
\end{equation}

\noindent In the above, $\mu$, $\nu$, and $\gamma$
are complex numbers, while $F=F^\dagger$ 
and $G=G^\dagger$ are operator--valued nonlinear 
functions of the fundamental quadrature operators
\begin{equation}
X_1 = \frac{a + a^\dagger}{2} \, , \; \; \; \; \;
\; \; \; \; \; X_2 = \frac{a - a^\dagger}{2i} \; .
\label{eq:Equaz3}
\end{equation}

The transformation is obviously a linear mapping 
in the space of operators. On the other hand,
it is in general a nonlinear, quadrature--dependent
extension of the 
Bogoliubov transformation in the sense that
the latter is
recovered either if $\gamma = \eta = 0$, or
if $F$ and $G$ contain only linear powers of the 
quadratures.

To begin with, we will study, for simplicity, only
the quadrature--dependent Bogoliubov transformation
generated by the first 
quadrature $X_1$ alone,
i.e. we put $\eta =0$ in Eqs.~(\ref{eq:Equaz1})
and (\ref{eq:Equaz2}) and thus consider the
transformation
\begin{equation}
b = \mu a + \nu a^\dagger + \gamma F(X_1) \; ,
\label{eq:Equaz4}
\end{equation}

\noindent together with its adjoint
\begin{equation}
b^\dagger = \mu^* a^\dagger + \nu^* a + 
\gamma^* F(X_1) \; .
\label{eq:Equaz5}
\end{equation}

In fact, all 
the main features of the quantum states generated 
by the quadrature--dependent Bogoliubov
transformations are already contained in this instance.
The extension to transformations generated by $G(X_{2})$ 
alone, and by both $F(X_{1})$ and $G(X_{2})$ will be
briefly discussed in the conclusions.

{\it i) Canonical structure}.
We first analyze under what conditions the
quadrature--dependent Bogoliubov transformation
(\ref{eq:Equaz4})--(\ref{eq:Equaz5}) is canonical. 
With a little 
algebra, we have that $[b, b^\dagger]=1$ if
\begin{eqnarray}
\label{eq:Equaz6}
|\mu|^{2} - |\nu|^{2} & = & 1 \; , \nonumber \\
&& \nonumber \\ 
{\mathrm{Re}} ( \mu\gamma^* - \nu^* \gamma ) & = & 0 \; , 
\end{eqnarray}

\noindent where ${\mathrm{Re}}(z)$ denotes the real part of 
the complex number $z$.
It is a remarkable feature that the conditions of
canonicity for the quadrature--dependent Bogoliubov
transformation are relations for the $c$--number
coefficients of the transformation
only, and do not involve the operator--valued
function $F(X_{1})$.
The same is true also when considering the 
quadrature--dependent Bogoliubov transformations
generated by $G(X_{2})$ alone, or by both $F(X_{1})$ 
and $G(X_{2})$.
Therefore, the
form of the operator--valued functions
$F(X_{1})$ and $G(X_{2})$ is
not constrained by requiring the transformation to
be canonical. 

The relations (\ref{eq:Equaz6}) are such
to give enough freedom in the choice of 
the coefficients of the transformation. This would not
be true if we had attempted to generalize the linear
Bogoliubov transformation introducing nonlinear
operator--valued
functions $A(a)$ and/or $B(a^\dagger)$ of the fundamental 
mode variables rather than of the quadratures. It is
easy to verify that in this case the conditions of
canonicity lead to such strict constraints
on the coefficients of the transformation
that, in practice, only the linear case $\gamma = 0$
is attainable. Thus it is the requirement of
canonicity that selects the quadratures
as the arguments of the nonlinear operator--valued
functions in the canonical transformations.

We see that the first of relations (\ref{eq:Equaz6})
is exactly the same as in the linear case ($\gamma = 0$),
but it is now constrained to be compatible with an
additional relation, the second of Eqs. (\ref{eq:Equaz6}),
which connects $\gamma$, the cofficient of the nonlinear part
of the transformation (\ref{eq:Equaz4})--(\ref{eq:Equaz5})
with the coefficients
$\mu$ and $\nu$ of the linear part.

{\it ii) Eigenstate of $b$: the multiphoton squeezed state.}
We now introduce, in strict analogy with the linear
case, the eigenstate $|\Psi\rangle_{\beta}$ of the transformed
annihilation operator $b$:
\begin{equation}
b|\Psi\rangle_{\beta} = \beta|\Psi\rangle_{\beta} \; ,
\label{eq:Equaz7}
\end{equation}

\noindent where $\beta$ is a $c$--number. The eigenvalue
equation (\ref{eq:Equaz7}) defines a new class of quantum 
states, the multphoton squeezed states,  
which are a direct generalization of the two--photon 
coherent states originally introduced by Yuen via the
linear Bogoliubov transformation. In principle,
depending on the choice of $F(X_1)$, the Bogoliubov
nonlinear squeezed states include 
all different generalized many--photon coherent
states, from four--photon coherent states on, as shall
be illustrated below.

\section{Hamiltonian structure}

As already mentioned, an important feature of the new 
quantum states
defined by Eq.~(\ref{eq:Equaz7}) is that, thanks to
the canonical quadrature--dependent Bogoliubov transformation
(\ref{eq:Equaz4})--(\ref{eq:Equaz5}), they can be associated
to a non quadratic Hamiltonian structure in terms of the
original mode variables $a$ and $a^\dagger$.
In fact, the quadratic Hamiltonian in the transformed variables 
\begin{equation}
H = b^\dagger b \; 
\label{eq:Equaz8}
\end{equation}

\noindent reads, in terms of the original mode variables,
\begin{eqnarray}
\label{eq:Equaz9}
H & = & (|\mu|^2 + |\nu|^2)a^\dagger a + 
\mu^* \nu a^{\dagger 2} + \mu \nu^* a^2 + |\nu|^2 +
\frac{1}{2} \nonumber \\
&& \nonumber \\
&+& \mu^* \gamma a^\dagger F(X_1) +\mu \gamma^*
F(X_1)a + \nu \gamma^* F(X_1)a^\dagger +\nu^*
\gamma a F(X_1) +|\gamma|^2 F^2 (X_1) \; . 
\end{eqnarray}

The first line of the Hamiltonian (\ref{eq:Equaz9})
contains the original Yuen squeezing Hamiltonian, 
including the two--photon 
contributions $a^{\dagger 2}$ and $a^2$. The remaining
terms appearing in the second line introduce anharmonicities,
as they involve higher powers
of $a$ and $a^\dagger$, whose degree depends
on the form of $F(X_1)$.
For instance, if we choose the lowest nonlinear 
power, i.e. $F(X_1) = X_1^2$, additional linear, 
quadratic, cubic, 
and quartic terms in $a$ and $a^\dagger$ are generated.
We are thus dealing with the effective description
of a single--mode quantum system undergoing  
multiphoton down conversion processes (parametric
down conversion processes of higher
order). The full Hamiltonian in the case $F(X_1) =
X_1^2$ reads
\begin{eqnarray}
\label{eq:Equaz10}
H & = & \frac{|\gamma|^2}{4} \left( a^{\dagger 4} +
a^4 \right) + |\gamma|^2 \left( a^{\dagger 3}a +
a^{\dagger}a^3 \right) + \frac{1}{4}|\gamma|^2a^{\dagger 2}
a^2 \nonumber \\
&& \nonumber \\
&+& \frac{1}{2}\left( \mu^* \gamma + \nu \gamma^* \right)
a^{\dagger 3} + \frac{1}{2}\left( \mu \gamma^* + \nu^* \gamma
\right) a^3 + {\mathrm{Re}} \left[ (\mu + \nu)\gamma^* \right] 
\left( a^{\dagger 2}a + a^{\dagger}a^2 \right) \nonumber \\
&& \nonumber \\
&+& \left( \frac{|\gamma|^2}{2} + \mu^* \nu \right) a^{\dagger 2}
+ \left( \frac{|\gamma|^2}{2} + \mu \nu^* \right) a^2 +
\left( |\mu|^2 + |\nu|^2 + 2|\gamma|^2 \right) a^{\dagger}a 
\nonumber \\
&& \nonumber \\
&+& \frac{1}{2} \left( \mu^* \gamma + 3\nu\gamma^* +
2\nu^* \gamma \right) a^{\dagger} + \frac{1}{2} \left(
\mu\gamma^* + 2\nu\gamma^* + 3\nu^* \gamma \right) a +
\frac{|\gamma|^2}{4} + |\nu|^2 + \frac{1}{2} \; . 
\end{eqnarray}

\noindent In general, choosing $F(X_1) = X_1^n $ with $n \geq 2$
will generate a $2n$--photon Hamiltonian. However, if we think
of $\gamma$ as the strength of the effective coupling to the
optical medium with higher nonlinearity, we see from 
Eqs.~(\ref{eq:Equaz9})--(\ref{eq:Equaz10})
that in the typical weak--coupling situations that are
realistically foreseeable the dominant terms will be those
associated to the lowest powers of $\gamma$. One can of
course consider trascendental forms for $F(X_1)$, such as, e.g.,
$F=\sin{X_1}$; also in these cases we expect that in the 
weak--coupling regimes the lowest terms in a power series
expansion of $F$ will contain the dominant contribution from
the nonlinear part of the Hamiltonian.

It is instructive to write the Hamiltonian
(\ref{eq:Equaz9}) in terms of
the quadratures $X_1$ and $X_2$, as this clarifies the
physical effects associated to the different terms.
We adopt, due to canonicity, the standard parametrization
$\mu = \cosh{r}e^{i\phi_1}$, $\nu = \sinh{r}e^{i\phi_2}$.
We have (letting $\gamma = |\gamma|e^{i\delta}$):
\begin{eqnarray}
\label{eq:Equaz11}
H &=& \left[ e^{2r}(1+\cos{(\phi_1 -\phi_2 )})
+ e^{-2r}(1 - \cos{(\phi_1 -\phi_2 )})\right] \frac{X_1^2}{4}
+ \left[ e^{2r}(1-\cos{(\phi_1 -\phi_2)}) +
e^{-2r}(1+\cos{(\phi_1 -\phi_2)}) \right] 
\frac{X_2^2}{4} \nonumber \\
&& \nonumber \\
&+& \sinh{2r}\sin{(\phi_1 -\phi_2)}\left\{ X_1 ,X_2 \right\} 
+ 2\sqrt{2}|\gamma|\cosh{r}\cos{(\phi_1 -\delta)}
X_1 \cdot F(X_1) \nonumber \\
&& \nonumber \\
&-& \frac{|\gamma|}{\sqrt{2}} \left[
\cosh{r}\sin{(\phi_1 -\delta)} - \sinh{r}\sin{(\phi_2 
-\delta)} \right] \left\{ F(X_1), X_{2} \right\} +
|\gamma|^2 F^{2}(X_1) \; ,
\end{eqnarray}

\noindent where $\{ \, , \, \}$ denotes the anticommutator.
We see from Eq.~(\ref{eq:Equaz11}) that the Hamiltonian
contains the standard terms associated to squeezing;
however the additional terms generated by $\gamma F(X_1)$
clearly modify the distribution in phase space with respect
to the standard two--photon squeezing, and this in turn
affects both the structure of the states and the observable
properties, as will be seen below. In fact, the 
quadrature--dependent Bogoliubov transformation
really amounts to a drastic deformation of the second
quadrature $X_{2}$, i.e. the momentum in particle language.
This can be most easily seen by considering the particular
case $\phi_1 = \phi_2 = \phi$. In such a situation, due to
the canonical constraints (\ref{eq:Equaz6}) the phase
$\phi$ of the coefficients $\mu$ and $\nu$ and the phase
$\delta$ of the coefficient $\gamma$ must obey the relation 
$\phi = \delta + ((2k+1)\pi)/2$, and the transformed
Hamiltonian reduces to
\begin{equation}
H = \frac{e^{2r}}{2}X_1^2 + \frac{e^{-2r}}{2}\left( X_{2} 
+ 2|\gamma|e^{r}F(X_1) \right)^{2} \; .
\label{eq:Equaz12}
\end{equation}

\noindent 
The above expression shows immediately that in the limit
$\gamma \rightarrow 0$ the Hamiltonian (\ref{eq:Equaz12})
reduces exactly to the standard squeezing Hamiltonian.
We thus see that the canonical quadrature--dependent 
Bogoliubov transformation is an effective description
of the coupling
of the system to an external medium with the following
overall effects: a squeezing transformation in both quadratures 
and a shift in the second quadrature (the canonical momentum in 
particle picture, the field phase in optical picture). The latter
is enforced by the nonlinear function $F(X_1)$, and for $F=X_1$
it reduces to the well--known Yuen shift in the momentum.
Of course, the shift $F(X_1)$ is not and must not be 
interpreted as an external electromagnetic vector potential. 
Instead, it summarizes the effect of the interaction with the 
external medium.
In turn, the shift affects the squeezing properties according
to the specific form of $F(X_1)$ and modifies the structure of
the stable, closed trajectories in phase space. We thus speak
of quadrature--dependent squeezing and of quadrature--dependent
shift and distorsion (deformation) of phase--space trajectories
and phase--space distributions. However, before turning to a
detailed analysis of the statistical properties (photon number
distribution and correlation functions) and of the phase--space 
distributions we first discuss the structure
of the uncertainty products and of the eigenstates of the 
transformation.

\section{Uncertainty products and asymptotic minimum uncertainty}

Let us consider the quantities 
$\Delta^2 X_i = \langle X_i^2 \rangle -
\langle X_i \rangle^2 $ $(i=1,2)$.
We want to evaluate the uncertainty product 
$\Delta^2 X_1 \Delta^2 X_2$ in the multiphoton
squeezed state  
$|\Psi \rangle_{\beta}$ solution of the eigenvalue equation 
(\ref{eq:Equaz7}). In terms of the transformed modes
$b$ and $b^\dagger$ the fundamental quadratures read
\begin{eqnarray}
\label{eq:Equaz13}
X_1 &=& \frac{1}{2}\left[ (\mu -\nu )^*b + 
(\mu-\nu)b^{\dagger} \right] \; , \nonumber \\
&& \nonumber \\
X_2 &=& X_2^{(l)} + X_2^{(nl)} \; ,
\end{eqnarray}

\noindent where 
\begin{eqnarray}
\label{eq:Equaz14}
X_2^{(l)} &=& \frac{i}{2}\left[ (\mu +\nu)b^\dagger -
(\mu + \nu)^*b \right] \; , \nonumber \\
&& \nonumber \\
X_2^{(nl)} &=& {\mathrm{Im}} (\mu\gamma^* -\nu^*\gamma)
F\left[ \frac{(\mu -\nu)^*b + (\mu-\nu)b^\dagger}{2}
\right] \; ,
\end{eqnarray}

\noindent where ${\mathrm{Im}} (z)$ stands for 
the imaginary part of the complex number $z$.
The expression for the first quadrature $X_1$
is identical to the one obtained via the linear 
Bogoliubov transformation: this is obviously due 
to the fact that the nonlinear part of the canonical 
quadrature--dependent Bogoliubov transformation
depends only on $X_1$.
What is changed with respect to the linear case
is evidently the expression for the second quadrature
$X_2$. We can see that it is made of two terms: the
first term $X_2^{(l)}$ is identical to that obtained
through the linear Bogoliubov transformation, while
the second term $X_2^{(nl)}$ is the contribution due
to the nonlinear part of the transformation, 
where the argument $X_1$
of the nonlinear operator--valued function $F$ 
has been re--expressed 
using the the first of Eqs.~(\ref{eq:Equaz13}).

Direct evaluation of the uncertainties in the state
$|\Psi \rangle_{\beta}$ yields for the first quadrature
\begin{equation}
\Delta^2 X_1 = \frac{1}{4}|\mu -\nu|^{2} \; ,
\label{eq:Equaz15}
\end{equation}

\noindent which is obviously still the same expression
originally obtained by Yuen in the linear case.
For the second quadrature one obtains
\begin{equation}
\Delta^2 X_2 = \Delta^2 X_2^{(l)} +
\Delta^2 X_2^{(nl)} + \langle 
\left\{ X_2^{(l)},X_2^{(nl)} \right\} \rangle
- 2\langle X_2^{(l)}\rangle \langle X_2^{(nl)}
\rangle \, .
\label{eq:Equaz16}
\end{equation}

We see that the uncertainty in the second 
quadrature is substantially modified with
respect to the linear case: the canonical
quadrature--dependent Bogoliubov transformation introduces
additional terms; in particular, the presence of the
anticommutator
expresses the existence of correlations between the
linear and the nonlinear part of $X_2$. Obviously, 
the term $\Delta^2 X_2^{(l)}$ acquires the
same value as in the linear case:
\begin{equation}
\Delta^2 X_2^{(l)} = \frac{1}{4}|\mu+\nu|^{2} \; ,
\label{eq:Equaz17} 
\end{equation}

\noindent while, introducing 
$\langle \left\{ X_2^{(l)}, X_2^{(nl)} 
\right\} \rangle_{T} \equiv
\langle \left\{ 
X_2^{(l)},X_2^{(nl)} \right\} \rangle
- 2\langle X_2^{(l)}\rangle 
\langle X_2^{(nl)} \rangle$, the remaining
terms in Eq.~(\ref{eq:Equaz16}) read
\begin{eqnarray}
\label{eq:Equaz18}
\Delta^2 X_2^{(nl)} & = & {\mathrm{Im}}^2(\mu\gamma^* -\nu^*\gamma)
\Delta^2 F(X_1) \; , \nonumber \\
&& \nonumber \\
\langle \left\{ X_2^{(l)}, X_2^{(nl)} \right\} 
\rangle_{T} & = & {\mathrm{Im}}(\mu\gamma^* -\nu^*\gamma)
\langle \left\{ X_2^{(l)}, F(X_1) \right\}
\rangle_{T} \; ,
\end{eqnarray}

\noindent where $\langle \left\{ X_2^{(l)}, F(X_1) \right\}
\rangle_{T} = \langle \left\{ X_2^{(l)}, F(X_1) \right\}
\rangle -2\langle X_2^{(l)}\rangle \langle F(X_1)
\rangle$. At variance with the uncertainty (\ref{eq:Equaz17})
in $X_2^{(l)}$, the remaining terms (\ref{eq:Equaz18})
that contribute to the overall uncertainty in $X_2$ 
depend explicitely on the eigenstate $|\Psi \rangle_{\beta}$
through the eigenvalue $\beta$.
Collecting terms together, the uncertainty product reads
\begin{eqnarray}
\label{eq:Equaz19}
\Delta^2 X_1 \Delta^2 X_2 &=& \frac{1}{16}|\mu^2-\nu^2|^2
+\frac{1}{4}{\mathrm{Im}}^2(\mu\gamma^*-\nu^*\gamma)|\mu-\nu|^2
\Delta^2F(X_1) \nonumber \\
&& \nonumber \\
&+& \frac{1}{4} {\mathrm{Im}} (\mu\gamma^*-\nu^*\gamma)|\mu-\nu|^2
\langle\left\{X_2^{(l)},F(X_1)\right\}\rangle_{T} \; .
\end{eqnarray}

Setting $\gamma=0$ in Eq.~(\ref{eq:Equaz19}) and choosing
$\phi_1=\phi_2=\phi$ in the expressions for the coefficients
$\mu$ and $\nu$ of the linear part of the transformation,
the uncertainty product reduces to the standard Heisenberg 
minimum with equal and opposite squeezing in the quadratures
$X_1$ and $X_2$. In the general case $\gamma \neq 0$ the
extra terms are in general non--zero. In the particular
case $\mu$ and $\nu$ both real or with equal phases, then 
$\Delta^2 X_1 = (1/4)\exp{(-2r)}$, 
$\Delta^2 X_2^{(l)} = (1/4)\exp{(2r)}$. 
Moreover, in this case, 
letting $\gamma = \gamma_1 + i\gamma_2$,
the conditions of canonicity (\ref{eq:Equaz6}) imply 
${\mathrm{Im}} (\mu\gamma^*-\nu^*\gamma)
=-(\gamma_2\exp{(r)})/\cos{\phi}$.
If we choose $F(X_1)$ of the form $F=X_1^n$ with 
the integer $n \geq 2$, then $\Delta^2 F \simeq \exp{(-2nr)}$
and $\langle\left\{X_2^{(l)},F(X_1)\right\}\rangle_{T} \simeq
\exp{(-2(n-1)r)}$. Collecting terms together we finally
have
\begin{equation}
\Delta^2 X_1 \Delta^2 X_2 \simeq \frac{1}{16}
+ \gamma_{2}^{2}e^{\textstyle -2nr} 
+ \gamma_{2}e^{\textstyle -2nr} \; .
\label{eq:Equaz20}
\end{equation}

We thus see that for large values of $r$ and/or for
small values of $\gamma_2$ the uncertainty product
is close to the Heisenberg minimum, and we can speak
of states of quasi minimum uncertainty. 

\section{Solution of the eigenvalue equation in 
the quadrature representation}

{\it i) Multiphoton squeezed vacuum}. We first consider
the equation for the vacuum state $|\Psi\rangle_{0}$
of the transformed mode $b$, i.e. for the multiphoton
squeezed vacuum:
\begin{equation}
b|\Psi\rangle_{0} = 0 \; .
\label{eq:Equaz21}
\end{equation}

\noindent In the quadrature ($X_1$) representation 
the operator $X_1$ acts as a multiplication by
the $c$--number $x$, and the equation for the
vacuum state reads
\begin{equation}
\left[ \left( \cosh{r}e^{\textstyle i\phi_1}
\right) \frac{x+iX_2}{2} +
\left( \sinh{r}e^{\textstyle i\phi_2}
\right) \frac{x-iX_2}{2} +
|\gamma|e^{\textstyle i\delta}F(x)\right]\Psi_{0}(x) 
= 0 \; ,
\label{eq:Equaz22}
\end{equation}

\noindent where the wave function
$\Psi_{0}(x)=\langle X_1|\Psi\rangle_{0}$ is the
multiphoton squeezed vacuum
in the quadrature representation.
We need to recall that the conditions of canonicity 
(\ref{eq:Equaz6}) impose
\begin{equation}
\cosh{r}\cos{(\phi_1-\delta)} -
\sinh{r}\cos{(\phi_2-\delta)} = 0 \; .
\label{eq:Equaz23}
\end{equation}

Multiplying both sides of Eq.~(\ref{eq:Equaz22}) by
$\exp{(-i\delta)}$, exploiting relation (\ref{eq:Equaz23})
and solving for $\Psi_{0}(x)$ we finally obtain
\begin{equation}
\Psi_{0}(x) = {\cal{N}}_{0}^{-1/2}
\exp{\left( -\frac{ \textstyle x^2}{ \textstyle 2\sigma} 
\right)}\exp{ \left[ 
\frac{i}{C_{-}} \left( \cosh{r}\cos{(\phi_1-\delta)}x^2 +
2|\gamma|\int^{x}dx'F(x') \right) \right] } \; ,
\label{eq:Equaz24}
\end{equation}

\noindent where ${\cal{N}}_{0}$ is the normalization
factor, $\sigma = C_{-}/C_{+}$ is the variance
of the Gaussian density profile of the wave function
and
\begin{eqnarray}
\label{eq:Equaz25}
C_{+} &=& \cosh{r}\sin{(\phi_1-\delta)}
+ \sinh{r}\sin{(\phi_2-\delta)} \; , \nonumber \\
&& \nonumber \\
C_{-} &=& \cosh{r}\sin{(\phi_1-\delta)}
- \sinh{r}\sin{(\phi_2-\delta)} \; .
\end{eqnarray}

The structure of the squeezed vacuum is thus the following:
the probability density is of Gaussian shape centered in 
$\langle X_1 \rangle =0$, while the function
$F(x)$ enters only in the phase of the wave
function. The phase is made up by two contributions:
a typical squeezing correlation term proportional to
$x^2$ plus a term proportional to the integral of 
$F(x)$. As the parameters of the transformation, and 
consequently the wave function, may
in principle be time--dependent, one can easily
derive the equations of motion for the center of
the wave packet $\langle X_1 \rangle(t)$
and for the squeezing parameter $r(t)$.
Considering the case $\phi_1=\phi_2=\phi$ one has
$\phi=\delta \pm ((2k+1)\pi)/2$, i.e. 
$\cos{(\phi-\delta)}=0$, $C_{+}=\exp{(r)}$,
$C_{-}=\exp{(-r)}$, $\sigma=\exp{(-2r)}$, 
and the wave function reduces to a Gaussian
modulated by a phase factor which includes
only the integral of the function $F(x)$.

{\it ii) Multiphoton squeezed states}.
We turn now to the solution of the eigenvalue
equation (\ref{eq:Equaz7}) that defines the
multiphoton squeezed states
for arbitrary complex values of the
eigenvalue $\beta=|\beta|\exp{(i\xi)}$.
Introducing the wave function
$\Psi_{\beta}(x) = \langle X_1|\Psi\rangle_{\beta}$
the eigenvalue equation (\ref{eq:Equaz7}) is easily
solved in the general case along the same lines
exposed in the case of the multiphoton squeezed vacuum
and the solution reads
\begin{equation}
\Psi_{\beta}(x) = {\cal{N}}_{c}^{-1/2}
\exp{\left[ -\frac{\left(x-x_{0}\right)^{2}}{2\sigma}
\right]} \exp{\left[ -\frac{i}{C_{-}}
\left(\cosh{r}\cos{(\phi_1-\delta)}x^2
+ |\beta|\cos{(\xi-\delta)}x 
-2|\gamma|\int^{x}dx'F(x')\right)\right]} \; ,
\label{eq:Equaz26}
\end{equation}

\noindent where the normalization ${\cal{N}}_{c}^{-1}
= 1/\sqrt{\pi\sigma}$, and 
$x_{0}=(|\beta|/C_{+})\sin{(\xi-\delta)}$ is the
center of the Gaussian probability profile.
The phase of the wave function contains both
the $x$--dependent and the $x^2$--dependent terms
typical of the squeezed coherent states of the
harmonic oscillator, plus an additional anharmonic
contribution coming from the $F(x)$--dependent term. 
In the particular case
$\phi_1=\phi_2=0$ 
(i.e. $\delta = \pm ((2k+1)\pi)/2$),
introducing the
complex number $\alpha=\alpha_1+i\alpha_2$ we
can choose $\beta=\mu\alpha+\nu\alpha^*$,
which, for the harmonic oscillator, is the condition 
of equivalence between the Yuen two--photon coherent
states and the standard squeezed coherent states obtained
first by squeezing the vacuum and then by displacing
it. We thus have, writing $\beta=\beta_1 +i\beta_2$,
$\beta_1=\alpha_1\exp{(r)}$ for the real
part, and $\beta_2=\alpha_2\exp{(-r)}$
for the imaginary part. The wave function
then reduces to
\begin{equation}
\Psi_{\beta}(x) = {\cal{N}}_{c}^{-1/2}
\exp{\left[ -\frac{e^{\textstyle 2r}}{2}
\left(x-\sqrt{2}\alpha_1\right)^{2}
\right]} \exp{\left[i\left( \sqrt{2}\alpha_2x 
- e^{\textstyle r}|\gamma|
\int^{x}dx'F(x')\right)\right]} \; .
\label{eq:Equaz27}
\end{equation}

\noindent The state (\ref{eq:Equaz27}) describes
a coherent squeezed dynamics:  
$\langle X_1 \rangle = \sqrt{2}\alpha_1$ 
is still equal to the mean in a coherent state
$\Psi_{\alpha}(x)$ of the harmonic oscillator,
while the variance $\sigma=\exp{(-2r)}/2$ coincides
with the dispersion $\Delta^2  X_1$ of the Yuen
two--photon coherent state. 
The equation for the wave--packet center
is easily derived, and one obtains
\begin{equation}
\frac{d^2 \langle X_1 \rangle}{dt^2}
= -\langle X_1 \rangle + e^{-r} \left( \langle F(X_1)
\rangle \frac{d |\gamma|}{dt} + |\gamma| \langle
\partial_t F(X_1) \rangle \right) \; ,
\label{eq:Equaz28}
\end{equation}

\noindent which reduces to the equation
of motion for the classical harmonic oscillator
when neither $\gamma$ nor $F(X_1)$ are explicitly
time--dependent. 
Therefore the
multiphoton squeezed states
preserve some of the basic properties of the
squeezed coherent states of the harmonic oscillator.
However, the presence of the $F$--dependent term in the 
phase of the wave function has relevant effects on
other physical quantities, 
in particular, as shall be seen in the next sections,
on the photon statistics, the correlation properties
and the structure of the quasiprobability distributions
in phase space. We also note from the
above discussion that in all cases $F(x)$ 
enters only in the phase of 
the wave function: it is therefore always possible
to cast the Hamiltonian (\ref{eq:Equaz11}) in the
general form 
\begin{equation}
H=aX_1^2 +\left( bX_1 + cX_2 + \gamma F(X_1) 
\right)^{2} \; ,
\label{eq:Equaz29}
\end{equation}

\noindent with coefficients $a$, $b$, $c$
that depend on the parameters $\mu$, $\nu$,
$\gamma$ of the canonical quadrature--dependent
Bogoliubov transformation. 

\section{Field Statistics}

Let us consider the photon number distribution (PND)
in a multiphoton squeezed state: 
$P(n)=|\langle n|\Psi\rangle_{\beta}|^{2}$.
Due to the nonlinear nature of the function $F(X_1)$
it is in general impossible to write a closed
analytic expression for $P(n)$, which can however
be easily plotted numerically. To gain immediate
insight on the effect of the $F$--dependent term
that enters in the phase of the state
$\Psi_{\beta}(x)$
we have confronted the PND of the standard Yuen
two--photon coherent state with the PND of the
multiphoton squeezed state 
with the lowest nonlinear behavior $F(x)=x^2$. 
We compare at $\phi_1=\phi_2=0$, $\beta$ real, and
set the coupling $\gamma$ at intermediate small values.
In Fig.~1 we have plotted the PND of the multiphoton
squeezed state at 
$\gamma=0.1$, $\beta_1=3$ and squeezing parameter
$r=0.8$ versus the PND of the corresponding Yuen 
two--photon coherent state ($\gamma=0$). 
\begin{figure}
\begin{center}
\includegraphics*[width=9cm]{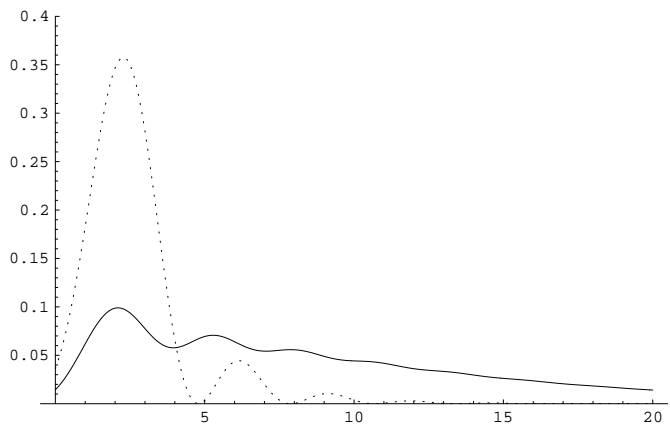}
\end{center}
\caption{The PND $P(n)$ of the multiphoton squeezed state
(\ref{eq:Equaz27}) 
(full line) plotted versus the PND of the corresponding 
two--photon coherent state (dotted line). For both PND's
the eigenvalue $\beta_1=3$ and the squeezing parameter
$r=0.8$. The PND of the multiphoton squeezed state
has been evaluated for
the lowest nonlinearity $F(x)=x^2$ and a coupling 
$\gamma=0.1$.}\label{fig01}
\end{figure}

We notice from Fig.~1
that the PND of the multiphoton squeezed state
lowers the 
first maximum that appears in the PND of the corresponding
two--photon coherent state at a very low number of photons.
Correspondingly, at a higher number of photons the PND of
the multiphoton squeezed state
stays well above zero approximately up to $n=20$ 
and well above the PND of the two--photon coherent state,
which is practically zero already at $n=10$. It is also
to be noted that the presence of the nonlinearity tends
to attenuate the oscillations in the PND, and we have
tested that this effect becomes more pronounced for
increasing values of the coupling $\gamma$. 
In fact, Fig.~1 shows that the minima of the oscillations 
in the PND of a multiphoton squeezed state
stay always well above zero. 
The explanation of this phenomenon is that the additional 
cubic nonlinear term in the phase of the wave function yields 
faster oscillations that enter in competition with the slower
oscillations due to the remaining linear and quadratic 
squeezing terms entering in the phase.
The net result is a decrease in the maxima and a
corresponding increase in the minima of the oscillations.
The effect is obviously more pronounced if we consider
the PND of a multiphoton squeezed state
with higher nonlinearity, as shown in
Fig.~2 for the case $F(X_1)=X_1^3$.
\begin{figure}
\begin{center}
\includegraphics*[width=9cm]{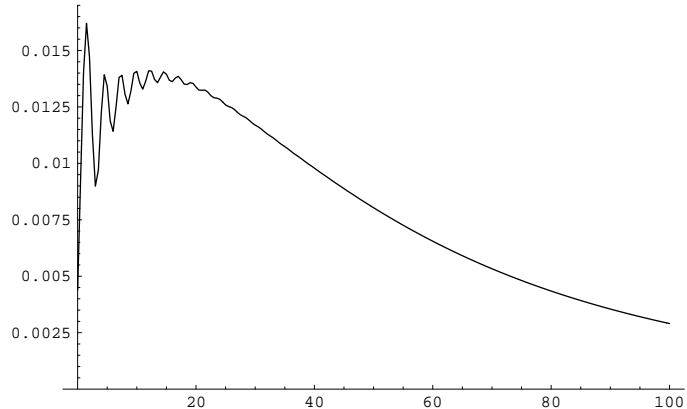}
\end{center}
\caption{The PND $P(n)$ of the multiphoton squeezed
state (\ref{eq:Equaz27}) 
plotted for a nonlinearity $F(X_1)=X_1^3$. As in the previous
case, the eigenvalue $\beta_1=3$, the squeezing parameter
$r=0.8$, and the coupling $\gamma=0.1$.}\label{fig02}
\end{figure}

A complementary explanation of this remarkable feature
in terms of areas of overlap between the Wigner functions
of number states and of multiphoton squeezed states
will be postponed to the section 
devoted to the phase--space analysis of the multiphoton
squeezed states.

Of special physical interest, among the expectations and 
the correlation functions of any order in the field 
operators, are the mean 
number of photons $\bar{n}=\langle a^\dagger a \rangle$
and the mean square deviation $\Delta^2 n =
\bar{n^{2}} - {\bar{n}}^{2} = \langle a^\dagger a
a^\dagger a \rangle - \langle a^\dagger a \rangle^{2}$
for the number operator $\hat{n} =a^\dagger a$.
The expressions of the first two moments of $\hat{n}$ are
rather complicated in the general multiphoton squeezed
state (\ref{eq:Equaz26}), but greatly simplify in the
multiphoton squeezed state (\ref{eq:Equaz27}).
In this state, choosing for instance 
with $F(X_1)=X_1^2$, $\phi=0$ (or $\delta = \pi/2$) 
and $\beta_2=0$, the mean photon number reads
\begin{equation}
\bar{n} = \frac{\left[ \left(2+3\gamma^2\right)\cosh{[2r]}
-3\gamma^2\sinh{[2r]}-2\right]}{4} + \beta_1^2\left(
\cosh{[2r]}-\sinh{[2r]}\right)\left(1+6\gamma^2+4\gamma^2
\beta_1^2\right) \; ,
\label{eq:Equaz30}
\end{equation}

\noindent which, 
reminding that $\beta_1^2=|\alpha|^2\exp{(2r)}$,
reduces to the known expression for the 
squeezed coherent states of the harmonic oscillator
$\bar{n}=\sinh^2{r}+|\alpha|^2$ when
$\gamma \rightarrow 0$.
The mean square deviation reads
\begin{eqnarray}
\label{eq:Equaz31}
\Delta^2 n &=& \frac{e^{-4r}}{8}\left[
\right. 1+e^{8r}+12\gamma^2+48\gamma^4+2e^{4r}
\left(\gamma^2-1\right) \nonumber \\
&& \nonumber \\
&+& 8\beta_1^2\left(1+\left(18+4e^{4r}\right)\gamma^2
+4\beta_1^2\gamma^2\left(4+e^{4r}+42\gamma^2+16\beta_1^2
\gamma^2\right)+96\gamma^4 \right) \left. \right] \; .
\end{eqnarray}

\noindent In the limit $\gamma \rightarrow 0$ 
the expression (\ref{eq:Equaz31}) reduces to
\begin{equation}
\Delta^2 n = \frac{e^{-4r}}{8}\left[ \left(
e^{4r}-1\right)^2+8|\alpha|^2e^{2r}\right] \; ,
\label{eq:Equaz32}
\end{equation}

\noindent which of course coincides with the
mean square deviation in the squeezed coherent
states of the harmonic oscillator. 
We see from the above expressions 
(\ref{eq:Equaz30})--(\ref{eq:Equaz31}) that
the multiphoton squeezed states can exhibit, 
just as in the linear
case, either sub--Poissonian or super--Poissonian
statistics depending on the assigned values
of the parameters.  To gain a deeper insight
of the field statistics it is useful to study
the second--order degree of coherence, i.e. the
normalized second--order correlation function
\begin{equation}
g^{(2)}(0) = \frac{\langle a^{\dagger 2}a^{2}
\rangle}{\langle a^{\dagger} a\rangle^{2}} =
1+\frac{\Delta^2 n -\bar{n}}{{\bar{n}}^2} \; ,
\label{eq:Equaz33}
\end{equation}

\noindent whose values allow to distinguish
between the different possible statistical
regimes. In particular, if $g^{(2)}(0) < 1$
the system exhibits sub--Poissonian statistics,
while for $g^{(2)}(0) > 1$ the statistics is
super--Poissonian.
In Fig.~3 the correlation function $g^{(2)}(0)$
of the squeezed coherent state of the harmonic
oscillator is plotted as a function of the parameter
of squeezing $r$ at $\beta_1=3$ 
(obviously here $\gamma=0$).
\begin{figure}
\begin{center}
\includegraphics*[width=9cm]{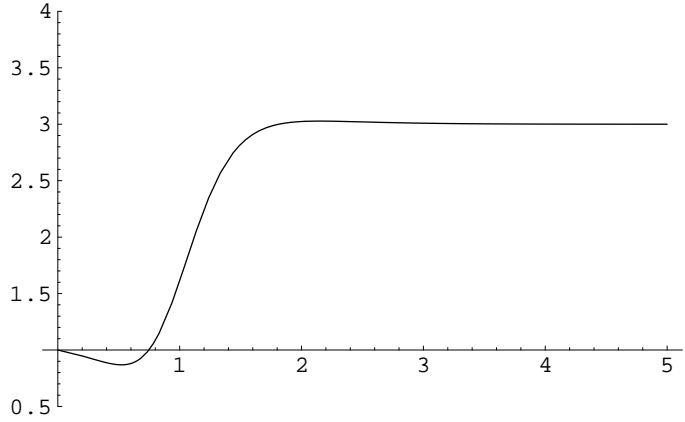}
\end{center}
\caption{The second--order correlation function $g^{(2)}(0)$ 
for the
multiphoton squeezed state
(\ref{eq:Equaz27})
plotted as a function of the 
squeezing parameter $r$ at $\gamma=0$ and $\beta_1=3$. 
It provides the statistical behavior of the harmonic 
oscillator squeezed coherent state. It is exactly
Poissonian ($g^{(2)}(0)=1$) at $r=0$ (coherent state). 
It then goes sub--Poissonian at small enough values of $r$, 
turns super--Poissonian at about $r=0.8$ until it saturates
at the value $g^{(2)}(0)=3$ for large values 
of $r$.}\label{fig03}
\end{figure}

Turning on the additional nonlinear Bogoliubov interaction with 
the external medium makes the correlation function $g^{(2)}(0)$
parametrically dependent on the coupling $\gamma$.
In Fig.~4 we show the second--order coherence plotted as
a function of $r$ again at $\beta_1=3$ but now with
$\gamma=0.05$ and quadratic nonlinearity $F(X_1)=X_1^2$.
\begin{figure}
\begin{center}
\includegraphics*[width=9cm]{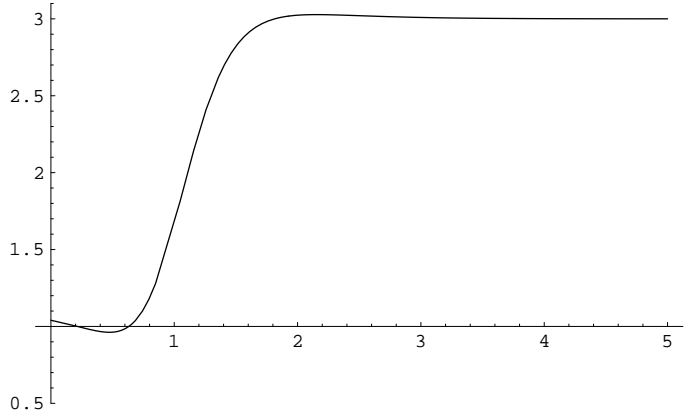} 
\end{center}
\caption{The second--order correlation function $g^{(2)}(0)$
for the multiphoton squeezed state (\ref{eq:Equaz27}) 
with quadratic nonlinearity $F(X_1)=X_1^2$
plotted as a function
of $r$ at $\beta_1=3$ and $\gamma=0.05$. The plot
shows that, at variance with 
the corresponding $g^{(2)}(0)$ for the squeezed coherent 
state of the harmonic oscillator ($\gamma=0$) shown in Fig.~3,
the multiphoton squeezed state
exhibits super--Poissonian statistics at $r=0$.}
\label{fig04}
\end{figure}

Fig.~4 shows that the statistical behavior is significantly
affected by the presence of the nonlinearity. Besides exhibiting
super--Poissonian behavior at $r=0$, the 
multiphoton squeezed state with $\gamma=0.05$
acquires a sub--Poissonian statistics in a more restricted range 
of values of $r$ (roughly from $r=0.2$ to $r=0.6$) compared to
the case $\gamma=0$. If we further increase the value of the
coupling $\gamma$, the multiphoton squeezed state
exhibits super--Poissonian statistics
for all values of $r$, including $r=0$, as shown in Fig.~5.
\begin{figure}
\begin{center}
\includegraphics*[width=9cm]{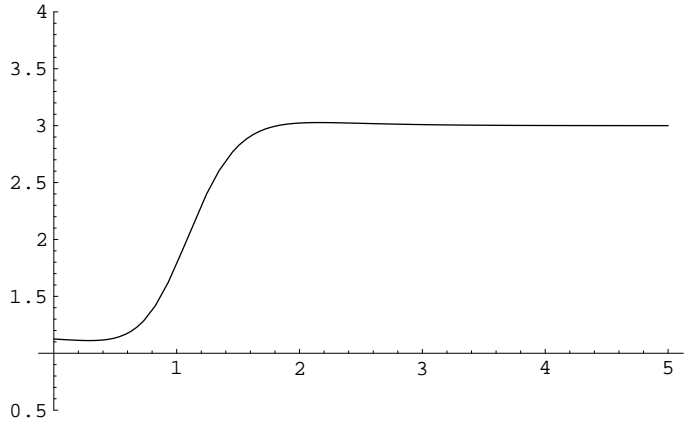} 
\end{center}
\caption{The second--order correlation function $g^{(2)}(0)$ 
for the
multiphoton squeezed state (\ref{eq:Equaz27}) 
with quadratic nonlinearity $F(X_1)=X_1^2$
plotted as a function of $r$ at
$\beta_1=3$ and $\gamma=0.1$. In this case the multiphoton
squeezed state 
exhibits super--Poissonian statistics for all values of
the squeezing parameter $r$.}\label{fig05}
\end{figure}

It is also interesting to study the field statistics
in a multiphoton squeezed state 
by tracking the behavior of the 
second--order coherence as a function of $\gamma$
for fixed values of the squeezing parameter $r$.
In Fig.~6 the correlation function $g^{(2)}(0)$ 
in the state (\ref{eq:Equaz27}) with quadratic 
nonlinearity $F(X_1)=X_1^2$ is 
studied as a function of $\gamma$ for $\beta_1=3$ and
$r=0.5$.
\begin{figure}
\begin{center}
\includegraphics*[width=9cm]{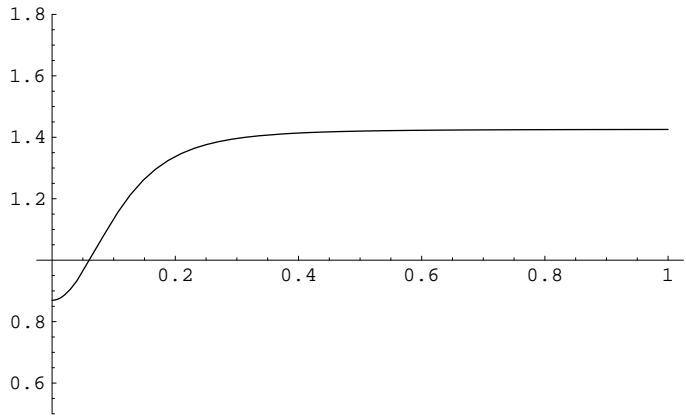}
\end{center}
\caption{The second--order correlation function 
$g^{(2)}(0)$ for the multiphoton squeezed state
(\ref{eq:Equaz27}) with quadratic nonlinearity
$F(X_1)=X_1^2$ plotted as a function of $\gamma$
at $\beta_1=3$ and $r=0.5$. The crossover between
sub-- and super--Poissonian behavior takes place
at about $\gamma=0.05$.}\label{fig06}
\end{figure}

The interplay between squeezing and nonlinear
deformation is seen most clearly by studying
the correlation function $g^{(2)}(0)$ at larger
values of the squeezing parameter $r$,
as shown in Fig.~7. For $r\geq 0.8$ the 
multiphoton squeezed state
(\ref{eq:Equaz27}) exhibits super--Poissonian
statistics for all values of $\gamma$.
\begin{figure}
\begin{center}
\includegraphics*[width=9cm]{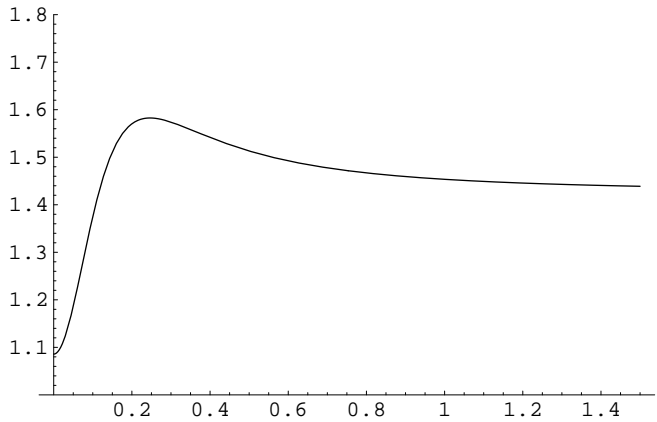}
\end{center}
\caption{The second--order correlation function 
$g^{(2)}(0)$ plotted as a function of $\gamma$
at $\beta_1=3$ and $r=0.8$. for the state 
(\ref{eq:Equaz27}) with quadratic nonlinearity
$F(X_1)=X_1^2$. The multiphoton squeezed states
exhibits in this case super--Poissonian statisitics
for all values of $\gamma$.}\label{fig07}
\end{figure}

\section{phase--space analysis}

The squeezed states of the harmonic oscillator are
typical nonclassical states of light. To study the
nonclassical features of the multiphoton squeezed states
and to compare them with
those of the squeezed states of the harmonic oscillator
it is most convenient to perform a phase--space analysis
in terms of the Wigner quasiprobability distribution
$W(X_1,X_2)$. It is well known that the Wigner function is 
positive--defined for the squeezed states of the harmonic
oscillator. Although well known, it is plotted
in Fig.~8 below with specific values of the parameters
for later comparison (We have relabelled the axes according
to ``particle'' language: $X_1 \equiv x$, $X_2 \equiv p$).
\begin{figure}
\begin{center}
\includegraphics*[width=9cm]{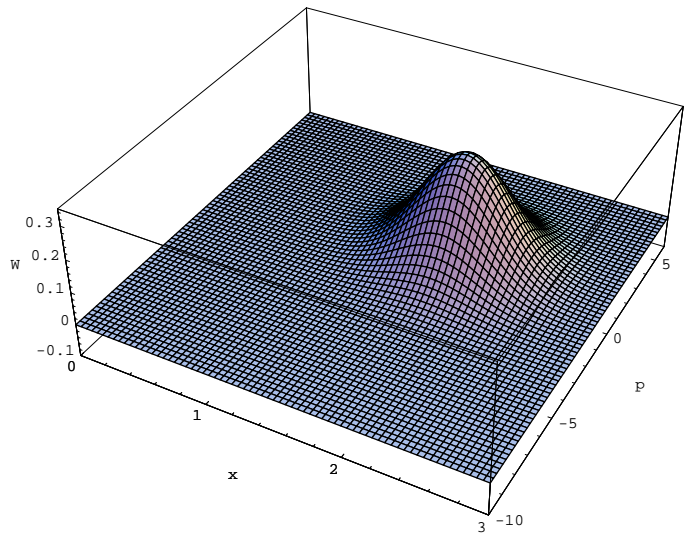}
\end{center}
\caption{The Wigner function of the harmonic oscillator
squeezed state, i.e. the multiphoton squeezed state
(\ref{eq:Equaz27}) with $\gamma=0$, at $\beta_1=3$, $r=0.8$.
The axes have been relabelled according to the ``particle''
language of phase space: 
$X_1 \equiv x$, $X_2 \equiv p$.}\label{fig08}
\end{figure}

The planar section of the same Wigner function is
also useful for later comparison, and it is
shown in Fig.~9 below. We notice that it is
centered at $X_2 \equiv p = 0$, and it is zero
outside a small interval $[-5,5]$ of variation
of $p$.
\begin{figure}
\begin{center}
\includegraphics*[width=9cm]{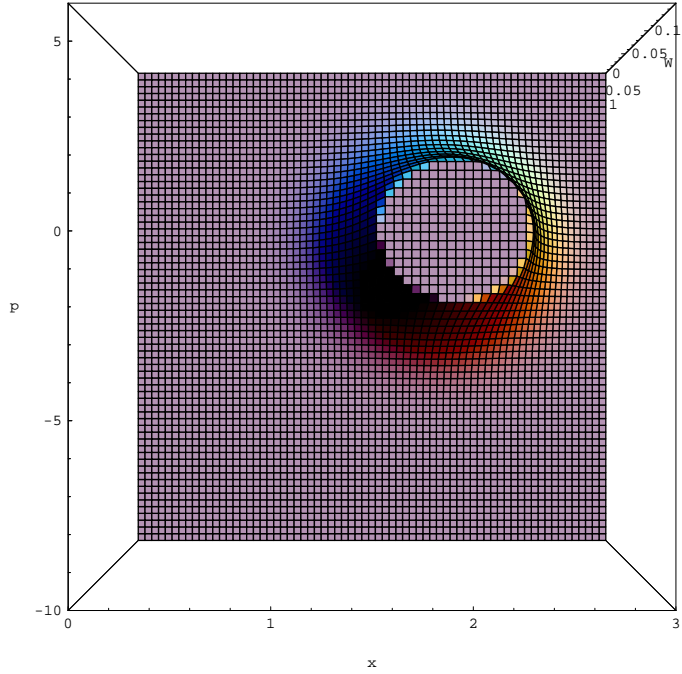}
\end{center}
\caption{The planar section of the Wigner
function for the harmonic oscillator
squeezed state, i.e. the multiphoton squeezed state
(\ref{eq:Equaz27}) with $\gamma=0$, at $\beta_1=3$, 
$r=0.8$.}\label{fig09}
\end{figure}

Positivity of the Wigner function is preserved
only by the multiphoton squeezed states with
quadratic nonlinearity $F(X_1)=X_1^2$. 
Even in this instance however, 
the Wigner function of the state (\ref{eq:Equaz27})
is displaced, rotated and deformed compared to
the corresponding distribution with $\gamma=0$, as 
shown in Fig.~10 below.
\begin{figure}
\begin{center}
\includegraphics*[width=9cm]{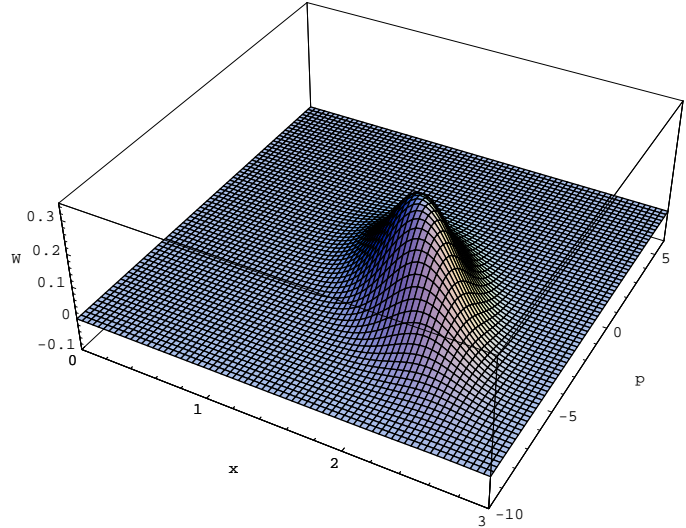}
\end{center}
\caption{The Wigner function of the multiphoton
squeezed state (\ref{eq:Equaz27})
with quadratic nonlinearity $F(X_1)=X_1^2$, 
at $\beta_1=3$, $r=0.8$, 
and $\gamma=0.1$. Notice that it is not centered 
in $p=0$, and it is elongated and rotated
compared to the purely squeezed case.}\label{fig10}
\end{figure}

More insight can
be gained by looking at its planar section. 
As shown in Fig.~11 the Wigner function of the
multiphoton squeezed state
exhibits an egg--shaped section (an elongated and
deformed ellipsis). Therefore
the area of overlap with the circular section 
associated to the generic number state of the
harmonic oscillator in  
phase space gives rise to a nontrivial interference which
is drastically modified with respect to the
purely squeezed case originally
discussed by Schleich, Walls and Wheeler \cite{schleich1},
\cite{schleich2}.   
This fact is responsible for the behavior of the oscillations 
in the photon number distribution of a multiphoton squeezed 
state (see Fig.~1).
\begin{figure}
\begin{center}
\includegraphics*[width=9cm]{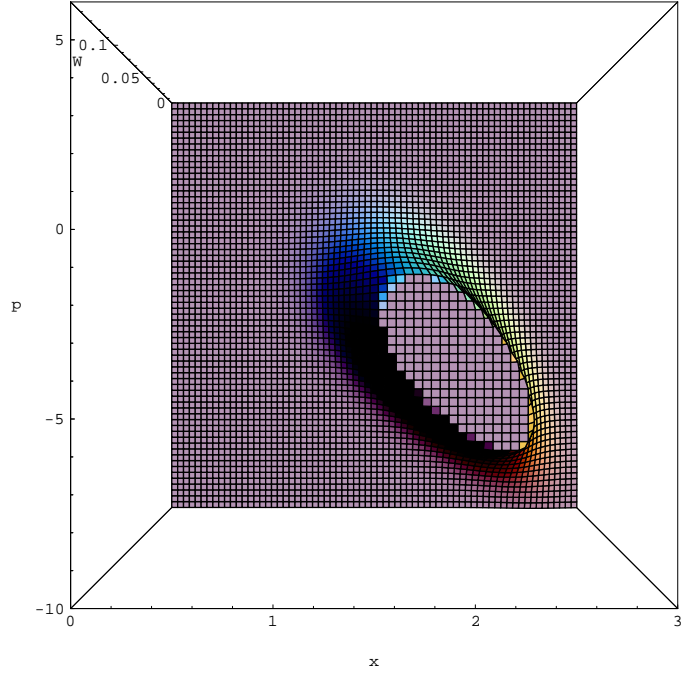}
\end{center}
\caption{Planar section of the Wigner function for 
the multiphoton squeezed state (\ref{eq:Equaz27})
with $F(X_1)=X_1^2$, $\beta_1=3$, $r=0.8$, and $\gamma=0.1$.
The elliptic section of the Wigner function for the
harmonic oscillator squeezed states is deformed into
a egg--shaped section.}\label{fig11}
\end{figure}

\noindent We see from Fig.~11 that besides the deformation
of the elliptic shape, the Wigner function
of the multiphoton squeezed state with quadratic
nonlinearity, although still positive defined, it is 
translated and rotated compared to the purely squeezed 
case. In particular, it is not any
more centered in $p=0$. Moreover it is elongated, and
it is non zero only in the range of negative values 
$[0, -10]$ of $p$.
The effects discussed above are significantly enhanced
if we consider multiphoton squeezed states
with higher nonlinearities.
Starting with the next higher nonlinearity
$F(X_1)=X_1^3$ the Wigner function of the
corresponding multiphoton squeezed state
becomes very
strongly deformed and acquires also negative values in
some regions of the phase space. 
The Wigner function for the multiphoton squeezed state
with cubic nonlinearity is reported in Fig.~12
below.
\begin{figure}
\begin{center}
\includegraphics*[width=9cm]{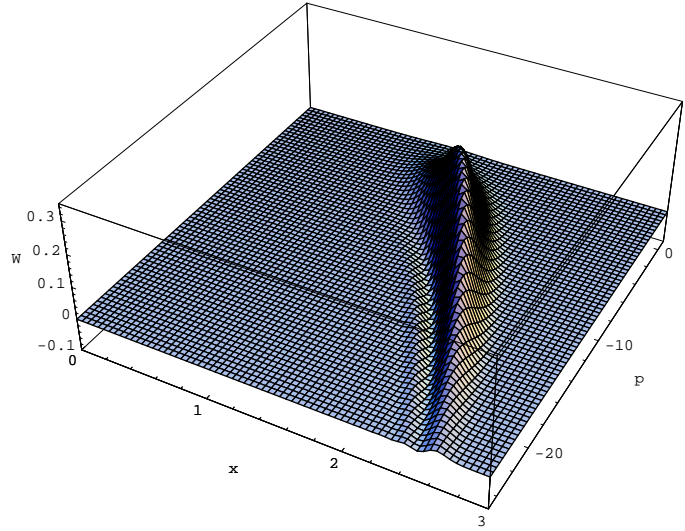}
\end{center}
\caption{The Wigner function for 
the multiphoton squeezed state (\ref{eq:Equaz27})
with cubic nonlinearity $F(X_1)=X_1^3$,
and the same values of the parameters $\beta_1=3$, $r=0.8$, 
and $\gamma=0.1$ as in the quadratic case $F(X_1)=X_1^2$.
Notice that it becomes negative along a strip of
values of $p$ ranging from $-2$ to $-24$ and of $x$
ranging from $1.4$ to $2.4$.}\label{fig12}
\end{figure}

\noindent We notice that, compared to the case with
quadratic nonlinearity, the Wigner function is more
rotated, tending to place itself parallel to the $p$
axis. It is also much more elongated, as it is non
zero for negative values of $p$ lying
in the range $[0, -25]$.
The planar section of the Wigner function 
for $F(X_1)=X_1^3$ is plotted in Fig.~13 below.
\begin{figure}
\begin{center}
\includegraphics*[width=9cm]{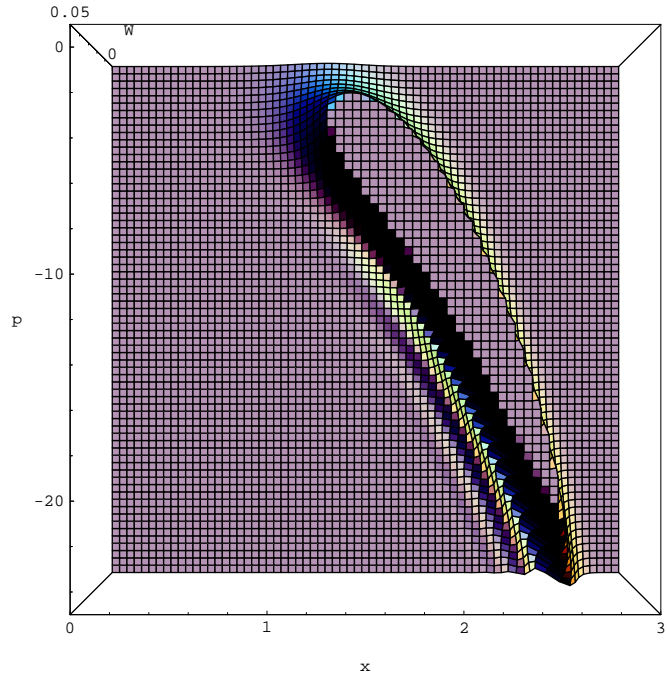}
\end{center}
\caption{Planar section of the Wigner function for 
the multiphoton squeezed state (\ref{eq:Equaz27})
with cubic nonlinearity $F(X_1)=X_1^3$,
at values $\beta_1=3$, $r=0.8$, 
and $\gamma=0.1$.}\label{fig13}
\end{figure}

\noindent We notice that the elliptic section becomes
completely deformed into a narrowing ``wing'' for
increasing negative values of $p$.  
Finally, in the case $F(X_1)=X_1^4$ the Wigner function becomes
even more twisted and shows many waves and ripples
with negative and positive peaks, as shown
in Fig.~14 below.
\begin{figure}
\begin{center}
\includegraphics*[width=9cm]{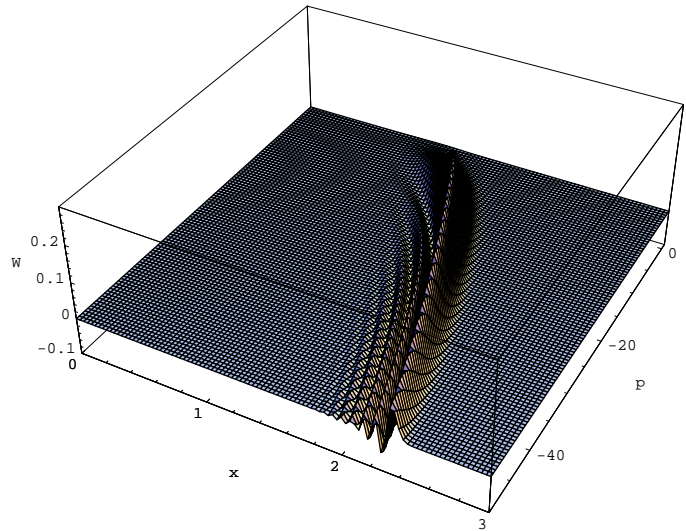}
\end{center}
\caption{The Wigner function for 
the multiphoton squeezed state (\ref{eq:Equaz27})
with quartic nonlinearity $F(X_1)=X_1^4$,
at values $\beta_1=3$, $r=0.8$, 
and $\gamma=0.1$.}\label{fig14}
\end{figure}

\noindent The planar section has the form reported in Fig.~15
below. It shows clearly the structure of the lateral waves
and ripples of negative and positive peaeks.
Besides, the central body is of still 
narrower ``wing'' section with further elongation, as the
Wigner function is now non zero in the interval $[0, -50]$
of negative values of $p$.
\begin{figure}
\begin{center}
\includegraphics*[width=9cm]{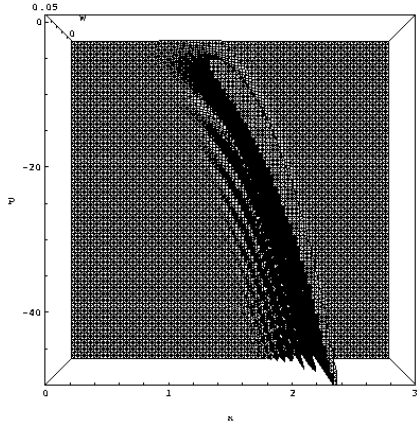}
\end{center}
\caption{Planar section of the Wigner function for 
the multiphoton squeezed state (\ref{eq:Equaz27})
with quartic nonlinearity $F(X_1)=X_1^4$,
at values $\beta_1=3$, $r=0.8$, 
and $\gamma=0.1$.}\label{fig15}
\end{figure}

\section{discussion and outlook}

In the present work we have introduced a new class of
nonclassical states of light. These new states have
been obtained via a quadrature--dependent generalization
of the linear Bogoliubov transformation first introduced 
by Yuen in his seminal work on the two--photon coherent
states of light. We have shown that our quadrature--dependent
Bogoliubov transformation, generated by a nonlinear operator
function of the first quadrature
is canonical under very broad constraints
on the numerical coefficients of the transformation.
The associated eigenstates of the transformation define
a new class of nonclassical states that preserve
some basic features of the coherent states of the harmonic
oscillator such as classical nondispersive motion of the
wave packet center. They also exhibit the amplitude squeezing
typical of the squeezed states of the harmonic oscillator.
However, the quadrature--dependent transformations generated
by quadratic and/or higher powers of the first quadrature
$X_1$ are associated to non quadratic effective
Hamiltonians that summarize the interaction with media
with higher nonlinearities (multiphoton down conversion
processes) and we thus name them as multiphoton squeezed
states. The presence of the quadrature--dependent 
term in the transformation is reflected in the
presence of an additional phase factor
in the wave function representation of these states.
The interplay of the additional phase factor with the standard
squeezing contributions is responsible for the remarkable
nonclassical behavior exhibited by the field statistics,
in particular by the photon number distribution and 
the second--order correlation function $g^{(2)}(0)$.
The phase--space analysis in terms of the Wigner 
quasiprobability distribution shows that the
multiphoton squeezed states are higly nonclassical,
as the Wigner function is strongly deformed
and it acquires negative values for cubic, quartic, and
higher nonlinearities. 

We have discussed the transformation generated by a 
nonlinear operator valued function of the first quadrature
$X_1$. It is of course possible to introduce the
transformations generated by a nonlinear operator
valued function of the second quadrature $X_2$ and
also by two nonlinear operator valued functions of
each quadrature. However the treatment is in these
cases slightly more tedious and involved, and will
be deferred to a follow--up of the present work.
In fact, looking at future developments, it
will be interesting to explore the possibility, that
we have only mentioned here, of considering 
quadrature--dependent transformations generated
by operator valued functions more general than
the simple powers considered in the present work.
For instance, considering the transformation
generated by a periodic function of the first
quadrature might allowe to define squeezed
states for massive particles interacting
with optical lattices.

We should finally mention the possibility, 
in contexts broader than 
quantum optics, of exploiting 
the quadrature--dependent Bogoliubov 
transformation as a technical tool for 
the introduction of normal modes. In particular,
we are currently investigating the possibility
of applying the quadrature--dependent Bogoliubov 
transformation to the approximate
diagonalization of the
Hamiltonian of the weakly interacting Bose gas.
The aim is to diagonalize a larger portion of
the Hamiltonian in the many--body Hilbert space
in comparison with the standard diagonalization 
originally performed via the linear Bogoliubov
transformation \cite{nozieres}.

\end{document}